\documentstyle[11pt,epsfig,a4wide]{article}
\newcommand{\mincir}{\raise
-2.truept\hbox{\rlap{\hbox{$\sim$}}\raise5.truept
\hbox{$<$}\ }}
\newcommand{\magcir}{\raise
-2.truept\hbox{\rlap{\hbox{$\sim$}}\raise5.truept
\hbox{$>$}\ }}
\newcommand{\minmag}{\raise-2.truept\hbox{\rlap{\hbox{$<$}}\raise
6.truept\hbox{$>$}\ }}
\newcommand{\be}{\begin{equation}}
\newcommand{\ee}{\end{equation}}
\newcommand{\ba}{\begin{eqnarray}}
\newcommand{\ea}{\end{eqnarray}}
\newcommand{\brr}{\begin{array}}

\newcommand{\err}{\end{array}}
\newcommand{\bc}{\begin{center}}
\newcommand{\ec}{\end{center}}

\title{Reconstruction of large-scale peculiar velocity fields}
\author{Roya Mohayaee$^1$, R.\ Brent Tully$^{1,2}$ and Uriel Frisch$^1$
\\
$^1$Observatoire 
de la C\^ote d'Azur, B.P.4229, F-06304 Nice Cedex 4, France\\
$^2$ Institute for Astronomy, University of Hawaii, Honolulu, HI 96822, USA\\
\\
{\sf Invited contribution to Colloquium, Cosmology: facts and Problems,}\\
{\sf Coll\`ege de France, Paris 8-11, June 2004}\\
{\sf Eds. J.V. Narlikar \& J-.C. Pecker}
}

\begin{document}

\maketitle

\begin{abstract}
A reconstruction method for recovering the initial conditions of the 
Universe starting from the present
galaxy distribution is presented which guarantees uniqueness of solutions. 
We show how our method can be used to obtain 
the peculiar velocities of a large number of galaxies,
hence trace galaxies orbits back in time and obtain the entire past dynamical
history of the Universe above scales where multi-streaming has not occurred.
When tested against a $128^3$ $\Lambda$CDM simulation in a box 
of $200$h$^{-1}$ Mpc length, we obtain $60\%$ exact reconstruction on
scales above $6$ h$^{-1}$Mpc. We apply our method 
to a real galaxy redshift catalogue, the updated NBG (Nearby Galaxies), containing 1483
galaxies, groups and clusters in a radius of $30$ Mpc/h, and reconstruct the
peculiar velocity fields in the local neighbourhood. Our reconstructed
distances are well-matched to the observed values outside the collapsed
regions if $\Omega_m(t)= 0.20\, \exp(-0.26 (t-13))$ where 
$t$ is the age of the Universe in Gyrs.

\end{abstract}

\section{Introduction}

Reconstruction of the initial condition of the Universe from the present distribution
of the galaxies, brought to us by ever-more sophisticated 
redshift surveys, is  
an instance of the general class of {\it inverse problems} in physics. 
In cosmology this problem is frequently tackled in an
empirical way by a {\it forward approach}. 
A {\it statistical} comparison between the outcome 
of an N-body simulation 
and the observational data is made, assuming that a suitable {\it bias} 
relation exists between the distribution of galaxies and that of dark 
matter. If the statistical test is satisfactory then the
implication is that the initial condition assumed by the simulation 
is a viable one for our Universe, otherwise one changes 
the cosmological parameters 
until a statistical convergence between the observed 
and the simulated present Universe  is achieved.

Since Newtonian gravity is time-reversible, one could integrate the equations of motions
back in time and solve the reconstruction problem trivially, if in 
addition to their positions, the present velocities of the 
galaxies were also known.
However, the peculiar velocities of only a 
few thousands of galaxies are known out of thousands whose 
redshifts have been measured. Thus, a second boundary condition, in addition
to the present redshifts of the galaxies, has to be provided :
as we go back in time the peculiar velocities of the galaxies vanish.
Contrary to the forward approach where one solves an {\it initial-value
problem}, in the reconstruction approach one is dealing 
with a {\it two-point boundary value problem} (In this case, only the
functional dependence of one of the boundary conditions is given, namely:
${\rm time}\rightarrow 0$ then ${\rm peculiar\,\, velocities}\rightarrow 0$). 
In the former, one has a {\it
unique} solution but in the latter this is not always the case. 

The question remains whether unique reconstruction 
can be achieved.
In this work, we report on a new method of reconstruction (Frisch et al. 2002,
Mohayaee et al. 2003, Brenier et al. 2003) which
guarantees uniqueness.

\section{A brief review of previous approaches to reconstruction}

The history of reconstruction goes back to the work of Peebles who
traced the orbits of the members of the Local Group back in time (Peebles
1989). In his approach, reconstruction was solved as a variational
problem. Instead of solving Newton's equations of motion, one searches
for the stationary points of the corresponding Euler-Lagrange action.
In his first work (Peebles 1989) only the 
minimum of the action was considered.
Later on, it was found that when the trajectories corresponding to the
saddle-point of the action were taken, a better 
agreement between predicted
and observed velocities could be obtained for the galaxies 
in the Local Group (peebles 1995). 
Thus, by adjusting the orbits until the 
predicted and observed velocities agreed, reasonable bounds on cosmological 
parameters were found (Peebles 1989) consistently favouring a low
density Universe ($\Omega_m=0.1-0.2$; noteworthy at a time when there was a
common preference for $\Omega_m=1$). 

Although rather successful (Shaya et al. 1995, Peebles et al. 2001) when applied to catalogues such as NBG
(Tully 1988) and also to mock catalogues (Branchini, Eldar\& Nusser 2002), 
reconstruction with such an aim, namely establishing
bounds on cosmological parameters using measured peculiar velocities, cannot be
applied to larger galaxy redshift surveys which contain
hundreds of thousands of galaxies for the majority of which the peculiar
velocities are unknown. For large
catalogues, the number of solutions become very large and not only
uniqueness is completely lost but also one is never sure that the full solution
space has been explored. In addition, numerical action-based 
codes are needed to solve the problem that challenges current computer capacities

A physical reason for multiple solutions is the
collisionless nature of cold dark matter. Collisionless fluid elements can undergo
multistreaming. Regions of multistream are bounded by {\it caustics} where the density is
formally infinite and inside which the velocity field can have more than one value.
This is a major obstacle to a unique reconstruction.

\section{Monge--Amp\`ere--Kantorovich (MAK) reconstruction}

Reconstruction can be a well-posed problem for as long as we avoid
multistream regions. The mathematical formulation of this problem is as
follows (see Frisch et al. 2002, Mohayaee et al. 2003 and Brenier et al. 2003). 
Unlike most of the previous works on reconstruction where one studies the
Euler-Lagrange action, we start from a constraint equation, namely the mass
conservation,
\be
\rho({\bf x})d{\bf x}=\rho_0({\bf q}) d{\bf q} \;,
\qquad
\ee
where  $\rho_0({\bf q})$ is the 
density at the initial position, ${\bf q}$, and 
$\rho({\bf x})$ is the density at the present position, ${\bf x}$, of the fluid
element. The above mass conservation equation can be rearranged in the
following form
\be
{\rm det}\left[{\partial q_i\over \partial x_j}\right]=
{\rho({\bf x})\over \rho_0({\bf q})}\;,
\label{det}
\ee
where ${\rm det}$ stands for determinant and $\rho_0({\bf q})$ 
is constant. The right-hand-side of the above
expression is basically given by our boundary conditions: 
the final positions of the particles are known
and the initial distribution is homogeneous, $\rho_0({\bf q}) ={\rm const}$. 
To solve the equation, we make the following hypotheses:
the Lagrangian map (${\bf q}\rightarrow{\bf x}$), is 
the {\it gradient} of a {\it convex}
potential $\Phi$. That is
\be
{\bf x}({\bf q},t)=\nabla_q\Phi({\bf q},t)\;.
\ee
The convexity guarantees that a single Lagrangian position corresponds to a 
single Eulerian position, {\it i.e.}, there has
been no multistreaming\footnote{ The gradient condition has been made in previous works
 (Bertschinger and Dekel 1989) on the reconstruction of the peculiar
 velocities of the galaxies using linear Lagrangian theory.}.
These assumptions imply that the inverse map ${\bf
 x}\rightarrow {\bf q}$ also has a potential representation
\be
{\bf q}={\bf \nabla}_{\bf x}\Theta ({\bf x},t)\;,
\ee
where the potential $\Theta({\bf x})$ is also a convex function
and is related to $\Phi({\bf x})$ by the Legendre--Fenchel transform
(e.g. Arnold 1978)
\ba
\Theta({\bf x})=\max_{\bf q}\left[
{\bf q}\cdot{\bf x}-\Phi({\bf q})\right]\,&;&\,
\Phi({\bf q})=
\max_{\bf x}\left[{\bf x}\cdot{\bf q}-\Theta ({\bf x})\right]\;.
\nonumber\\
\ea
The inverse map is now substituted in (\ref{det}) yielding
\be
{\rm det}\left[{\partial^2 \Theta({\bf x},t)
\over \partial x_i\partial x_j}\right]=
{\rho({\bf x})\over \rho_0({\bf q})}\;,
\label{ma}
\ee
which is the well-known Monge--Amp\`ere equation (Monge 1781, Amp\`ere 1820) 
The solution to this
222 years old problem has recently been discovered
(Brenier 1987, Benamou and Brenier 2000) when
it was realized that the map generated by the solution to the Monge--Amp\`ere
equation is the unique solution to an optimisation problem.
This is the Monge--Kantorovich mass transportation problem (Kantorovich 1942), in which one
seeks the map ${\bf x}\rightarrow {\bf q}$ which minimises the quadratic 
{\it cost} function
\be
I=\int_{\bf q} \rho_0({\bf q})\vert{\bf x}-{\bf q}\vert^2 d^3q=
\int_{\bf x} \rho({\bf x})\vert{\bf x}-{\bf q}\vert^2 d^3x \;. 
\label{cost}
\ee
A sketch of the proof is as follows. A small variation in 
the cost function yields
\be
\delta I=\int_{\bf x} \left[2\rho({\bf x})({\bf x}-{\bf q})
\cdot \delta {\bf x}\right] d^3x\;,
\ee
which must be supplemented by the condition
\be
{\bf\nabla}_{\bf x}\cdot\left(\rho({\bf x})\delta{\bf x}\right)=0\;,
\ee
which expresses the constraint that the Eulerian density remains unchanged. 
The vanishing of $\delta I$ should then hold for all ${\bf x}-{\bf q}$
which are orthogonal (in $L^2$) to functions of zero divergence. These are
clearly gradients. Hence ${\bf x}-{\bf q}({\bf x})$ and thus ${\bf q}(\bf x)$
is a gradient of a function of ${\bf x}$.

Discretising the cost (\ref{cost}) into equal mass units yields
\be
I=\min_{j(\cdot)}\left(\sum_{i=1}^N\left({\bf q}_{j(i)}-
{\bf x}_i\right)^2\right)\;.
\label{assign}
\ee
The formulation presented in (\ref{assign}) 
is known as the {\it assignment problem}: given $N$ initial and $N$ final
entries one has to find the permutation which minimizes the quadratic
cost function. The cost
function is indeed the minimum of an Euler-Lagrange 
action for inertial particles formulated in suitable space and time
coordinates (Croft \& Gazta\~naga 1997).
If one were to solve the assignment problem (\ref{assign}) for $N$ particles directly, 
one would need to search among $N!$ possible permutations, for the one which
would have the minimum cost. However, advanced assignment algorithms exist which
reduce the complexity of the problem from factorial to polynomial 
(e.g. see H\'enon 1995 and Bertsekas 1998. Furthermore H\'enon's adaptation of
sparse and dense algorithm suitable for cosmological problems has a complexity
of less than $N^{2.5}$ and has been used
extensively in Mathis, Mohayaee \& Silk 2004.)

\section{Test against numerical simulation}

\begin{figure}
\centerline{
        \epsfxsize=0.4\textwidth
        \epsfbox{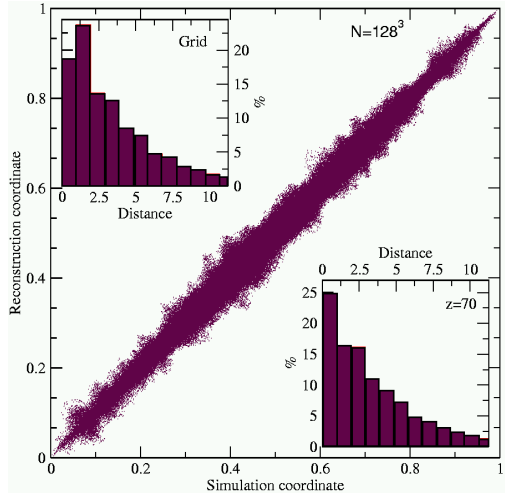}}
\vspace{0.05cm}
\caption
{
\small 
In the scatter plot, the dots near the diagonal
are a scatter plot of reconstructed initial points versus simulation initial
points for a grid of size $1.5$ Mpc/h with more than 2 million points. The scatter
diagram uses a {\it quasi-periodic projection} coordinate
${\tilde {\bf q}}\equiv (q_x+\sqrt 2 q_y+{\sqrt 3} q_z)/(1+
\sqrt 2+\sqrt 3)$ which guarantees a one-to-one correspondence between $\tilde
{\bf q}$ values and points on the regular Lagrangian grid. The upper left
inset is a histogram (by percentage) of distances in reconstruction mesh 
units between such points; the first bin corresponds to perfect
reconstruction;
the lower-inset is a similar histogram for reconstructed points at $z=70$.
The points at $z=70$ are obtained by using Zel'dovich approximation to push
particles back in time once their grid position has been reconstructed. 
Perfect reconstruction of about 18\% is achieved in both histograms
on scales of about $2$ Mpc.  On mesh sizes of about $6$
Mpc/h this rate increases to about 60\%.
}
\label{scattplushisto}
\end{figure}

We have tested our reconstruction against numerical N-body simulation. We ran
a $\Lambda$CDM simulation of $128^3$ dark matter particles, using the
adaptive P$^3$M code HYDRA (Couchman et al. 1995). Our cosmological 
parameters are $\Omega_m=0.3,
\Omega_\Lambda=0.7, h=0.65, \sigma_8=0.9$ and a box size of
$200$Mpc/h. The simulations started at high redshift, in this case at $z=70$. 
The results of our full box 
reconstruction are shown in Fig.\ \ref{scattplushisto}. 
Once the assignment problem is solved the peculiar velocities can be simply
evaluated using the zel'dovich approximation
$\dot{\bf x} = f(\Omega) H(t)\times({\bf x} - {\bf q})$
where $f(\Omega)=d{\rm ln}D/d{\rm ln} a$ is dimensionless linear growth rate,
$D(t)$ is the amplitude of the growing mode today, $a$ is the cosmic scale
factor and $H(t)$ is the value of the Hubble parameter (Zel'dovich 1970). The peculiar
velocities can then  be used to reconstruct the positions ${\bf x}$ of the
particles at any desired redshift back in time : ${\bf x}(z)={\bf
  q}+(D(z)/D_0)({\bf x}_0-{\bf q})$ where ${\bf x}_0$ is their present
positions, given by the simulation and  $D_0$ is the present value of $D$.
The lower-inset of Fig.\ \ref{scattplushisto} shows the exact rate of
reconstruction (when the separation between reconstructed and simulated
positions of the particles is less than one mesh at $z=70$) to be more or less
the same as that as the top left inset. The reason is that particles move very
little from the grid positions at high redshifts. However, a comparison
between the two histograms demonstrates that yet another Zel'dovich
approximation which is involved in getting from grid
positions to positions at $z=70$ does not
decrease the success of
our reconstruction. (For detailed tests against simulations and reconstruction of
statistics of the primordial density field, e.g. works on issues such as
non-Gaussianity, see Mathis, Mohayaee \& Silk 2004). For scales below $2$
Mpc corresponding to smallest scale probed by reconstruction whose results are
given in Fig.\ \ref{scattplushisto}, the exact reconstruction rate is about 18\%
due to severe multistreaming at these scales. On larger scales of about $5\%$
this rate increases to about 60\%. 

\begin{figure}
        \epsfxsize=0.9\textwidth
      {\epsfbox{velocities.eps}}
\ba
    \hspace*{-1.2cm}   \epsfxsize=0.39\textwidth \epsfbox{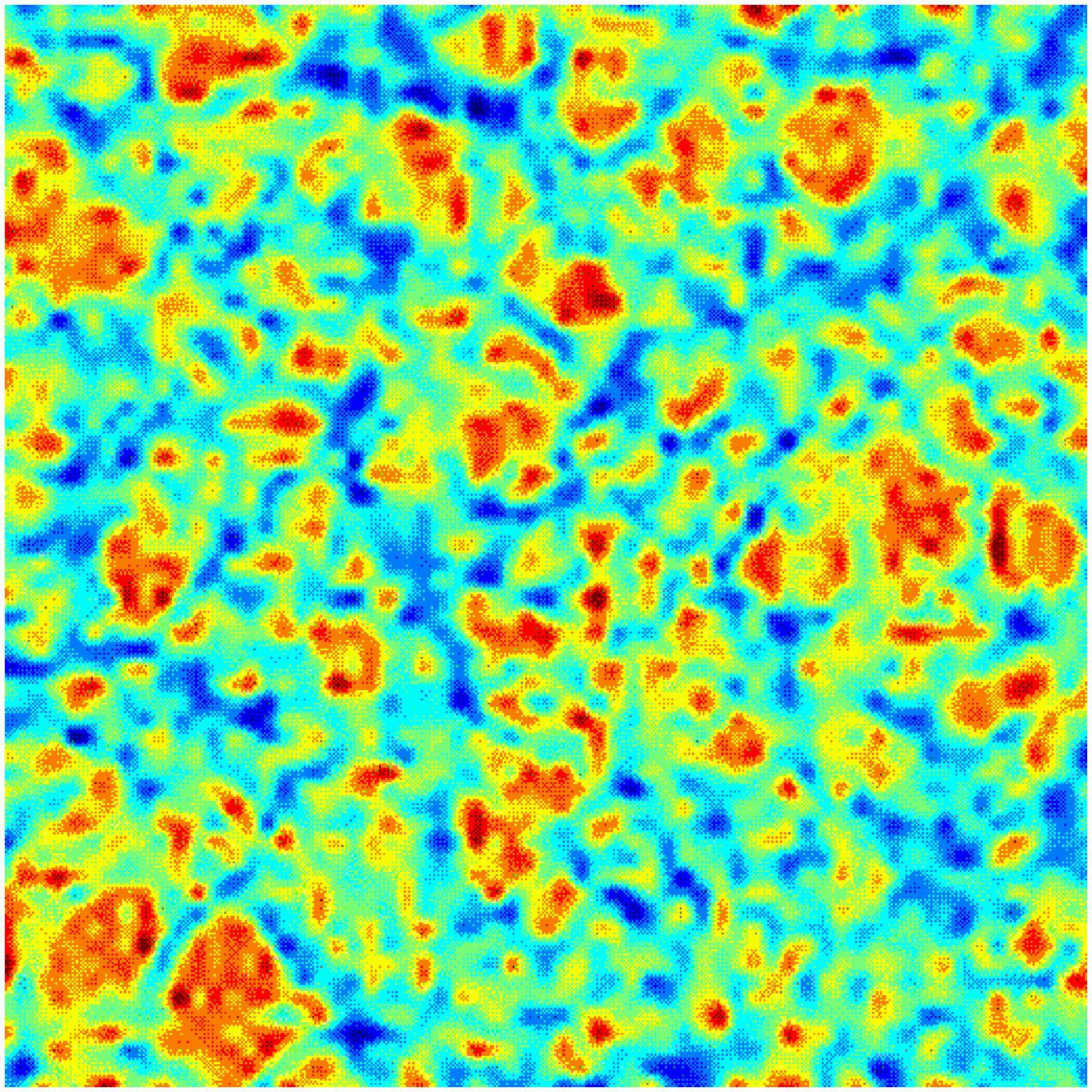}
&&
       \epsfxsize=0.39\textwidth \epsfbox{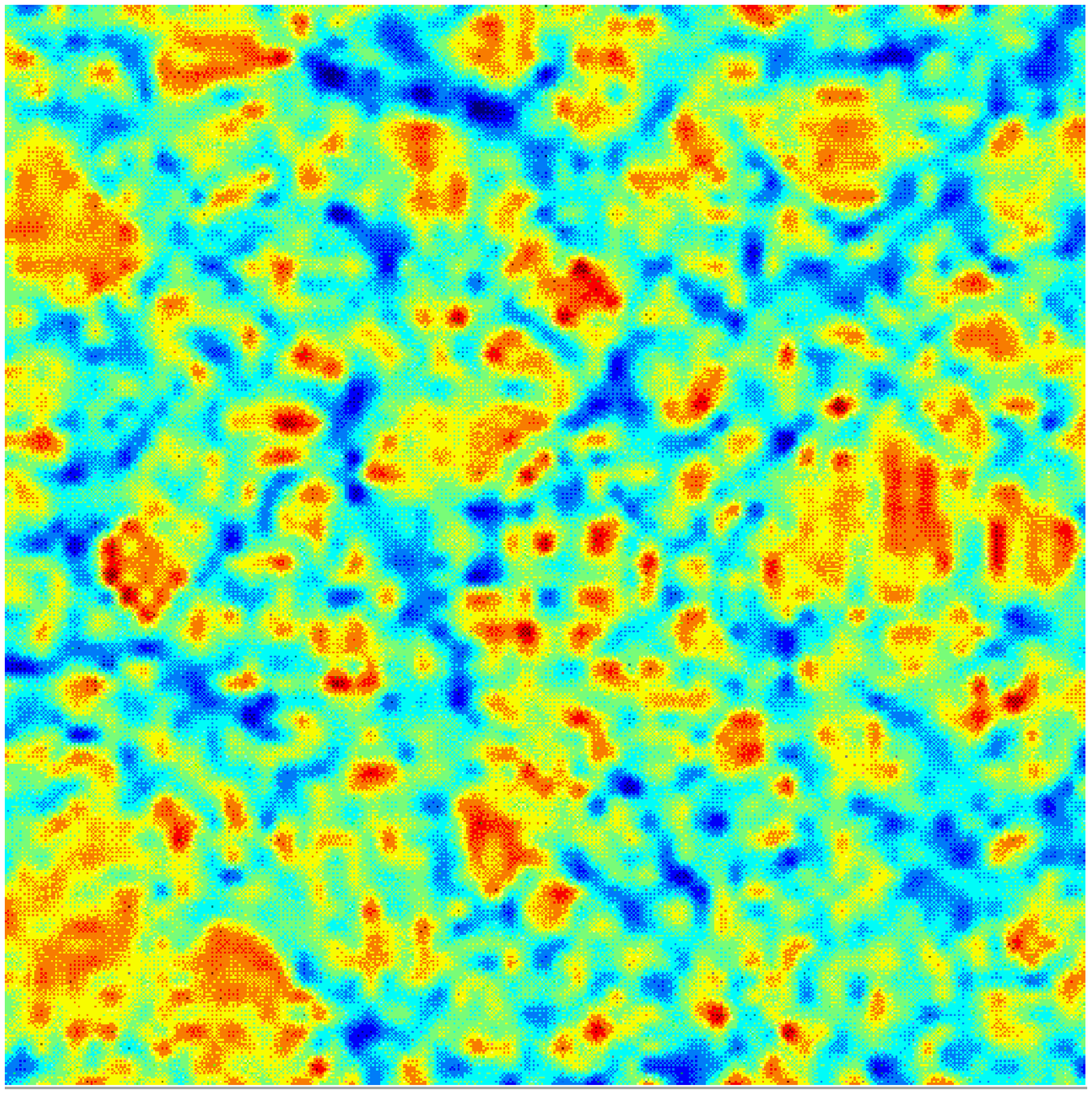}
\nonumber
\ea
\vspace{-0.2cm}           
\caption
{
Top panel:
simulated (left panel) and the reconstructed (right panel)
velocity field are shown for a thin 6 Mpc/h slice in $x$ and $y$ direction and
full-box projection in the $z$ direction of the simulation box. The
reconstruction works extremely well outside dense/collapsed regions.
bottom panel:
Simulated density field (left panel)
and the reconstructed (right panel) are shown for a thin slice cut of the
simulation box. Red colour corresponds
to denser regions and dark blue (lighter) are the 
void regions (cloud-in-cell interpolation is used). (Upper and lower
panels do not correspond to the same slices of the simulation box.)
}
\label{velocities}
\end{figure}

Outside collapsed regions the reconstructed peculiar velocities match well
those simulated as shown in Fig. \ref{velocities}. The primordial density
field evaluated using these velocities also matches extremely well the
simulated one as demonstrated in the 
lower panel of Fig.\ \ref{velocities} (we thank S. Colombi for
providing us with the lower panel of Fig.\ \ref{velocities}).

\section{Application to real galaxy catalogues}

We have applied our MAK method to the 
updated NBG catalogue (Tully 1988), now including 3300 galaxies within
3,000~km~s$^{-1}$.   Other more extensive catalogues are available but
this catalogue provides good completion within the specified volume,
which is sufficient in depth for present purposes.  The NBG has the
important value added feature of the detailed assignment of all
objects to homogeneously identified groups and filamentary structures.
The zone of Milky Way avoidance is shrinking as new surveys are integrated
but before a dynamical model can be computed something must be
done to account for galaxies lost due to obscuration.  In this work,
fake galaxies were created by reflection of objects at nearby higher
latitudes in sufficient numbers to achieve the average density for the
volume.  Another correction to the catalogue is one that accounts for
incompleteness with distance.  The correlation with mass is with the
quantity of blue light.  Light is lost from the catalogue as galaxies
become increasingly excluded with distance.  Fortunately the problem
is not extreme over the limited range of this study.  Selection
function corrections to luminosity range from unity at less than 10
Mpc (inside which there is completion because a
low luminosity clip is imposed at $M_B=-16$) to only a factor 2.4 at
3000 km/s.
The second observational component is a catalogue of galaxy distances.
In all, there are over 1400 galaxies with
distance measures within the 3,000~km~s$^{-1}$ volume.  
In the present study, distances are averaged over groups
because orbits cannot meaningfully be recovered on sub-group scales. 
The present NBG catalogue is assembled into 1234 groups (including groups
of one) of which 633 have measured distances.

This catalogue of galaxy positions, luminosities, and distances
provides the basis for orbit reconstructions using MAK procedures.  
The distances, $d$, permit an extraction of
peculiar velocities $V_{pec} = V_{gsr} - d {\rm H}_{0}$ where $V_{gsr}$ is
the observed velocity of an object in the galactic standard of rest.

\begin{figure}
\ba
\epsfxsize=0.4\textwidth\epsfbox{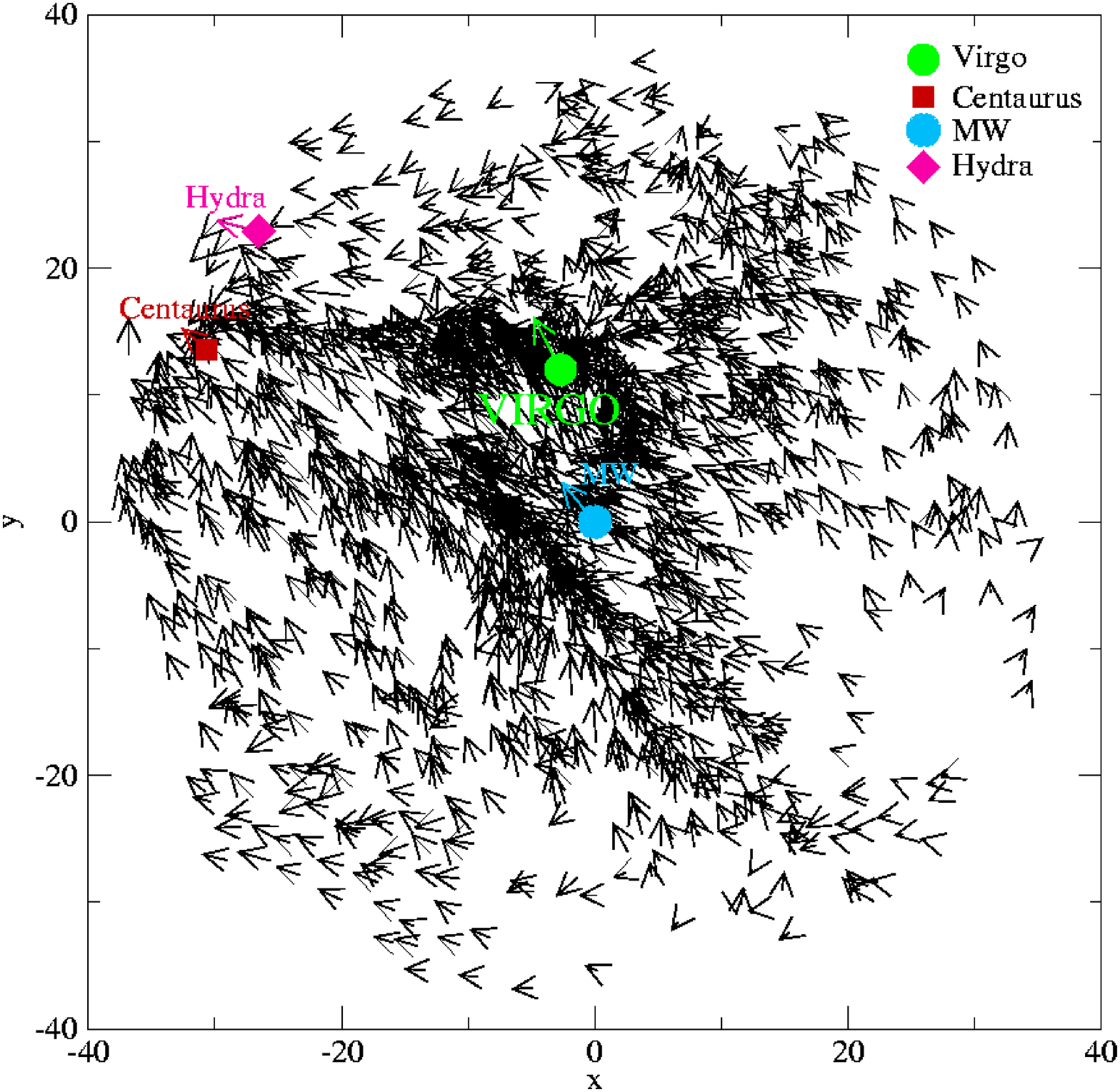}
&&
\epsfxsize=0.4\textwidth\epsfbox{ModulusMAK.eps}
\nonumber
\ea
\vspace{0.1cm}
\caption
{
Left plot: Velocity field of objects in NBG catalogue obtained by MAK reconstruction is
shown. Large-scale flow towards the great attractor is visible, which overshadows infall
into the Virgo cluster. The supergalactic coordinates $x$ and $y$ are
used. Right panel: is the scatter plot between MAK
reconstructed distance modulus and that given by observations for the $663$
objects with measured distances in this catalogue.
}
\label{nbgvelocity}
\end{figure}

For MAK reconstruction the particles must all have the same mass since all
the particles on the initial grid must be equal and each orbit
reconstruction has equal weight.  Consequently the endpoint elements
must be broken up by different amounts depending on their supposed
relative masses.  In the first approximation of constant mass-to-light ratio $M/L$ then
the elements are simply broken into a number of particles that depends
on $L_i$.  The unit size is chosen to correspond to $10^9 L_{\odot}$,
the faint end cutoff of the catalogue.
The elements are all located in redshift space ({\it i.e.}, at their positions
on the sky and at a distance inferred from their velocities).  However
the breakup into particles for the MAK reconstruction does not
preserve the velocity distortion from real positions within elements;
{\it i. e.}, on sub-group scales.  As we have demonstrated in the previous section, 
the MAK reconstructions of N-body simulations
demonstrates good recovery of orbits on scales greater than
$5h^{-1}$ Mpc but clearly orbits cannot be 
recovered in shell-crossing regions.  

The orbit of an element is
defined by the center of mass of all the constituent particles as a
function of time.  The relationship between redshift and real space is
estimated using the Zel'dovich approximation 
${\bf v} = f(\Omega)({\bf x} - {\bf q})$ where ${\bf v}$ is the peculiar
velocity vector, ${\bf x}$ is the current Eulerian position, ${\bf q}$
is the initial Lagrangian position, and $f(\Omega) \sim 1+b/\Omega_{m,0}^{4/7}
+(1+\Omega_{m,0}/2)\Omega_{\Lambda,0}/70$ and $b$ is the bias factor which
we take equal to $1$. In principle with our methodology, a variable bias can
be obtained by varying $M/L$ with location. In this discussion, the same $M/L$
is assigned to all objects.

Once the particles are reconstituted into the catalogue
elements, a specific model defines positions that can be tested against
observed positions. In Figures \ref{nbgvelocity}, we show two MAK results. The
left panel is the peculiar velocity field reconstructed by MAK of all the
entries in the NBG catalog. There is a clear flow towards the great attractor
as expected. The right panel shows a scatter plot of reconstructed versus
observed distance moduli,
$
\mu_i = 5{\rm log}d_i+25
$.
The scatter is mainly due to poor reconstructions near big clusters
such as Virgo. In the infall region of Virgo, one is in the highly non-linear
regime and moreover in a triple-valued region due to redshift space
distortion. In this region, velocities deviate significantly from Hubble flow
and MAK reconstruction does not necessarily find the right solution. (For
reconstruction in the infall region, used for determination of mass of Virgo
cluster, see Tully \& Mohayaee 2004.) 

The overall MAK reconstruction can be evaluated by a $\chi^2$ estimator.
We evaluate the median value for the $\chi_i^2$; between measured and observed
distance moduli
\be
\chi^2_i=(\mu_{observed}- \mu_{MAK})^2/\epsilon_i
\label{chisquared}
\ee
where $\epsilon_i$ is the error assigned to $\mu_{observed}$ which is the observed
distance modulus of galaxy (or group or cluster) $i$ in the catalogue.
Values of $\chi_i^2$ can be determined for the 633 objects in the catalogue with
distance measures for a given choice of density parameter $\Omega_m$ and age $t$.  

In this study we have only considered flat topologies.  We assume that
$\Omega_{\Lambda} = 1 - \Omega_m$ where $\Omega_{\Lambda}$ is a measure of the
energy density of the Universe.  With this constraint, there is a fixed
relation between $\Omega_m$, H$_0$, and the age of the Universe, $t$, such that if
two of these parameters are specified then the third is defined:
$h=(1/{\rm t})(2/3)
(1/\sqrt{(1-\Omega_m)}){\rm
  log}((1+\sqrt{(1-\Omega_m)})/\sqrt{(\Omega_m)})9.78$
where $h=H/100$.

\begin{figure}
\begin{center}
{
 \epsfxsize=0.6\textwidth  \epsfbox{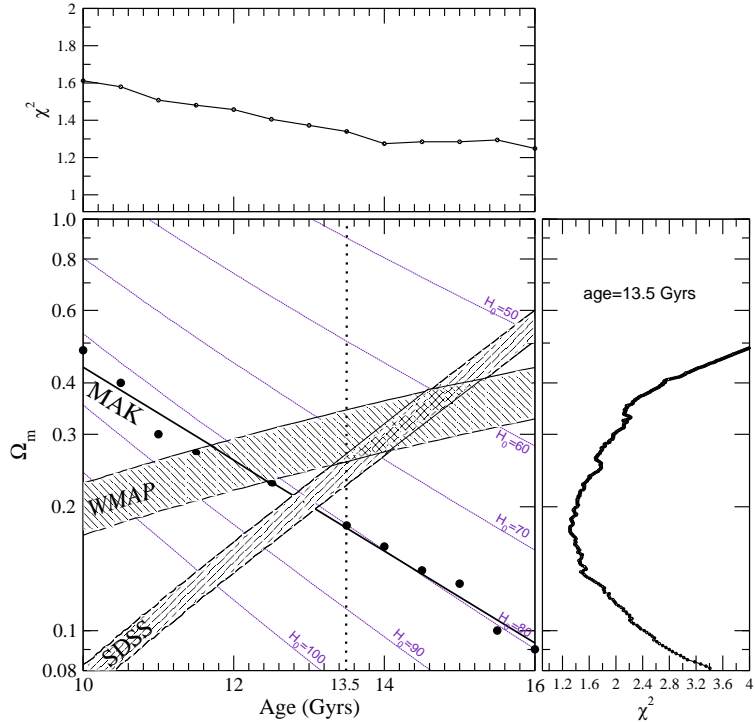}
}
\end{center}
\vspace{-0.1cm}
\caption
{
Constraints on the parameters $\Omega_m$ and age of the Universe given by MAK,
WMAP, and SDSS.   
The $2 \sigma$ constraints from WMAP and SDSS are given as
shaded bands.  The minimum of the $\chi^2$ trough with the MAK reconstruction
is given as the heavy solid line.   Hubble constant contours are superposed as
light lines.  The two side panels illustrate aspects of $\chi^2$ with the MAK
reconstruction.  In the top panel, the minimum value of $\chi^2$ is shown at
each age (ie, at the location of the heavy solid line).  In the right side
panel, the values of $\chi^2$ are shown for the range of $\Omega_m$ considered
for the specific age $t=13.5$ Gyr (ie, the trace indicated by the vertical
dotted line).  It is seen that there is reasonable agreement between the
three methodologies in the vicinity of $t=12-14$ Gyrs, $h=0.8$ and
$\Omega_m=0.2-0.3$. 
}
\label{parameters}
\end{figure}

Constraints on the parameter space ($\Omega_m, t$) are summarized in
Fig. \ref{parameters}.
The two broad bands locate the 95\% confidence limits provided by WMAP spatial
fluctuation and SDSS power spectrum studies (Spergel et al. 2003; Tegmark et
al. 2004).  The heavy solid line indicates the locus of $\chi^2$ minima as a
function of age from the MAK reconstructions.  This line is described by the
equation $\Omega_m(t)= 0.20\, \exp(-0.26 (t-13))$ with age $t$ in Gyr.
The right panel illustrates the
dependence of $\chi^2$ values on $\Omega_m$ at the fixed age of $t=13.5$ Gyr.
The top panel shows the weak dependence of $\chi^2$ on age at the
$\chi^2$-minimum trough defined by the heavy solid line.  The overall minimum
along this trough is reached at 17 Gyr.  Overall with Fig. \ref{parameters}
two important
points are to be noted.  First, the uncertainties resulting from the MAK
analysis are almost orthogonal to the WMAP and (especially) the SDSS
constraints.  Second, the three results intersect, resulting in concordance
with the cosmological parameters $ t=13.2 \pm 0.8$ Gyr, $\Omega_m = 0.25 \pm
0.05$, and H$_0 = 77 \pm 5$.

In conclusion, we have demonstrated that our MAK reconstruction scheme
guarantees uniqueness on large scales and can be applied to large datasets
containing millions of objects. It is now being used with real data for the
reconstruction of 
large-scale velocity fields.  The method has been tested against numerical
simulations and been shown to recover the peculiar velocities of a large number
of galaxies with a high success rate (taking the simulation dark matter
particles to trace galaxies). 
We have also shown that MAK can be
applied to real data and reconstructed peculiar velocity fields in the Local
Supercluster.  The best reconstruction fits obey the relationship
$\Omega_m(t)= 0.20 \exp (-0.26 (t-13))$, where $t$ is the age of the Universe
in Gyrs. This fit intersects the {\it WMAP} and SDSS results within
their $2\sigma$ uncertainties in the range $t$: 13-13.5 Gyrs, 
whence $\Omega_m=0.2-0.3$.


Materials presented in Section 4, are parts of an ongoing collaboration with 
M.\ H\'enon, S.\ Colombi and H.\ Mathis. 
Materials presented in Section 5 are parts of
collaborations with
J.\ Peebles, S.\ Phleps and E.\ Shaya. We also thank J.\ Colin, S.\ Matarrese and 
A.\ Sobolevskii for discussions and comments.
R.M.\ is supported by a European Marie Curie fellowship HPMF-CT2002-01532.
B.T.\ is partially supported by the BQR program 
of the Observatoire de la C\^ote d'Azur.

\end{document}